\documentclass[conference]{IEEEtran}
\IEEEoverridecommandlockouts
\usepackage{cite}
\usepackage{amsmath,amssymb,amsfonts}
\usepackage{algorithmic}
\usepackage{graphicx}
\usepackage{textcomp}
\usepackage{xcolor}
\usepackage{verbatim}
\usepackage{subfigure}
\def\BibTeX{{\rm B\kern-.05em{\sc i\kern-.025em b}\kern-.08em
    T\kern-.1667em\lower.7ex\hbox{E}\kern-.125emX}}
 
\begin{document}

\title{Cyber Security issues and Blockchain-Deep Learning based solutions for UAV and Internet of Drones (FANETs)\\
}

 \author{\IEEEauthorblockN{Partha Protim Datta}
 \IEEEauthorblockA{\textit{School of Computing} \\
 \textit{University of North Florida}\\
Jacksonville, USA \\
n01490529@unf.edu}
}

\maketitle

\begin{abstract}

Safety-critical systems such as automated embedded or industrial systems have a strong dependency on the trustworthiness of data collection. As sensors are the critical component for those systems, it is imperative to address the attack resilience of sensors.

\end{abstract}



\section{Introduction}

Unmanned aerial vehicles (UAVs), cumulatively known as drones, have become a ubiquitous presence in various industries such as agriculture, logistics, and surveillance. With the advent of the Internet of Things (IoT), UAVs are increasingly being used as part of IoT-enabled systems, forming what is called the Internet of Drones (IoD). This IoD networks also known as a type of wireless mobile ad-hoc network or FANET. This networks consist of unmanned aerial vehicles (UAVs) that communicate with each other in a peer-to-peer fashion to accomplish their tasks. However, the integration of UAVs into these systems has led to various security challenges such as secure routing, secure communication, secure positioning jamming, eavesdropping, and spoofing attacks [1]. Other forms of attacks can also be seen and addressed i.e. DoS attacks [3], unauthorized access, and Sybil attacks [2], by Cyber-attacks on UAVs and their communication networks, can cause significant damage to critical infrastructure and pose a threat to public safety~\cite{YAACOUB2020100218}. To address these challenges, researchers have proposed various security solutions, including the use of blockchain [4],[7],[10] agent-based self-protective methods, cryptographic [18] and deep reinforcement learning [12], [13], [14]. In this survey paper, the state-of-the-art security mechanisms are analyzed for threats applicable for the UAV-IoD applications.

\section{Methods}
In this review paper, the discussion for the methods can be developed in four stages.

\subsection{Threat Models}
Several studies have proposed different threat models in the context of UAV networks. Lateef et al.[1] made a comprehensive study on gathering an in-depth understanding of the state-of-the-art technology trends in FANETs and security threats. A novel methodology was proposed to address the identified gaps and limitations in the literature. Based on their research, potential security threats in FANETs were identified and categorized into different types, including data confidentiality, integrity, availability, authentication, and authorization. Mathematical equations and models from the literature were utilized to analyze the impact and severity of these threats on the security of FANETs. The proposed methodology integrated multiple security measures, such as encryption, authentication, and intrusion detection, into a comprehensive framework that accounted for the unique characteristics and requirements of FANETs. The experimental design included setting up realistic simulation scenarios that mimicked the real-world FANET environment, considering factors such as network size, mobility patterns, and communication protocols. The analysis of the data included evaluating the performance metrics, such as packet delivery ratio, end-to-end delay, and energy consumption, to assess the effectiveness of the proposed methodology in securing FANETs. The results of the validation were used to establish the effectiveness of the proposed methodology and support the conclusions of the study.

Elias Ghribi et al. [3] addresses threats such as denial of service (DoS) attacks, replay attacks, and insider attacks. The problem is clearly defined, emphasizing the need for a secure communication approach that can ensure the confidentiality, integrity, and availability of data exchanged among UAVs in a decentralized and trustless environment. They have mentioned about a secure blockchain-based communication approach for UAV networks, which employed a permissioned blockchain to ensure secure communication and data sharing among UAVs. 

Another DoS attack scheme has been challenged by Khan et al.[13]. The authors addressed Denial of Service (DoS) and Distributed Denial of Service (DDoS) attacks in the Internet of Flying Things. They explained that these attacks can be launched by an attacker who floods the network with a large amount of traffic, overwhelming the network's resources and making it unavailable to legitimate users. They highlighted that such attacks can result in serious consequences, including loss of control of the drone, data theft, and even physical damage. Therefore, they proposed an experience-based deep learning algorithm to detect and prevent such attacks in IoFT/IoD networks. The derived algorithm is trained on previous attack data to improve its effectiveness and adaptability to new attack patterns[13].

Rupa et al provides some background on data tampering, data leakage, and data forgery- these three types of security attacks[4]. This paper discusses the security and privacy concerns associated with the use of unmanned aerial vehicles (UAVs) and proposes a blockchain-based solution to address these concerns.

Basudeb bera et al focused on blockchain application on the FANET. They addressed IoD attacks on [2] . The evaluation of the proposed scheme using simulations and demonstrate its effectiveness in mitigating various attack scenarios, such as unauthorized access, Sybil attacks, and denial-of-service attacks.

Joseanne Viana et al. proposes a novel deep attention recognition architecture for identifying attacks in 5G Unmanned Aerial Vehicle (UAV) scenarios[11]. To address the growing 5G security issues the authors highlight the security concerns of UAVs in 5G networks and the need for effective attack detection mechanisms. They argue that traditional intrusion detection methods may not be sufficient to address the dynamic and complex nature of attacks in UAV scenarios.

To address this challenge, the authors propose a deep attention recognition architecture that combines convolutional and recurrent neural networks to analyze the visual and textual data captured by UAVs in real-time. The proposed architecture incorporates an attention mechanism that focuses on the most informative regions of the input data to improve the accuracy of attack identification. The authors conducted an end-to-end evaluation of the proposed architecture using a dataset of simulated 5G UAV attack scenarios. The results show that the proposed architecture achieves a high accuracy rate of 97.89\% in detecting attacks, outperforming other state-of-the-art approaches.

\subsection{Security Solution Methods}
Several defense mechanisms have been proposed to enhance the security and privacy of UAV networks. In broad category we can see three types of security.

\subsubsection{Blockchain based security}
Rupa et al discuss a study on the security and privacy of UAV data using blockchain technology, which highlighted the potential of blockchain to ensure data integrity, confidentiality, and authenticity in UAV systems [4]. The proposed solution involves the use of smart contracts, which can be used to define access control policies for UAV data. The authors also discuss the use of blockchain consensus algorithms, such as proof of work (PoW) and proof of stake (PoS), to ensure the integrity and authenticity of UAV data. They argue that the use of blockchain technology can enhance the security and privacy of UAV data by providing a tamper-proof and decentralized storage mechanism.  One limitation is the computational overhead associated with blockchain-based solutions, which can affect the real-time performance of UAVs. The authors suggest exploring the use of lightweight blockchain solutions to address this limitation. 

Another blockchain based method is proposed by Aparna et al[7]. The authors  introduces the concept of blockchain-enabled softwarization as a solution to these challenges. Blockchain-enabled softwarization refers to the use of blockchain technology to create a flexible and secure software-defined networking (SDN) framework for UAV networks.The paper presents a taxonomy of different mechanisms that can be used to enable blockchain-enabled softwarization in UAV networks. The taxonomy is based on four key components of blockchain-enabled softwarization: consensus algorithms, smart contracts, access control policies, and key management. For each component, the paper provides a detailed description and analysis of different mechanisms that can be used to enhance security and privacy in UAV networks.
The paper also provides a comprehensive review of related work in the area of blockchain-enabled softwarization for UAV networks. The review covers recent research on consensus algorithms, smart contracts, access control policies, and key management, and discusses the strengths and limitations of different approaches.

The use of blockchain has been accessed by Basudeb Bera et al on a different and very interesting way in IoT [2] [10]. In this paper they merged the idea of securing IoT network by blockchain-based access control scheme.This access control scheme is designed to be scalable, secure, and efficient. It supports dynamic policy updates and flexible policy enforcement. The authors propose a consensus protocol that ensures the consistency and integrity of the policy database. The scheme also employs cryptographic techniques to protect the privacy of the drone operators and the policy database. 

Blockchain and deep learning both merged on 5G communication by Dai et al.The authors tried to enhance the performance and capabilities of 5G networks by leveraging the power of blockchain and deep reinforcement learning. The paper proposed a novel framework that integrates blockchain and deep reinforcement learning to enhance the intelligence, security, and efficiency of 5G networks. The methodology includes the deployment of intelligent agents, which utilize deep reinforcement learning techniques to learn from the network environment and make intelligent decisions. The agents also collaborate with each other to share information and collectively optimize the network performance. The blockchain technology is used to ensure the integrity and security of the network transactions and data, while also enabling decentralized control and management. Performance metrics, such as network throughput, latency, and packet loss, were measured and analyzed to assess the efficiency and effectiveness of the proposed methodology. Security metrics, such as attack detection rate, false positive rate, and system resilience, were also evaluated to assess the security of the network. Statistical analysis techniques, such as hypothesis testing and data visualization, were used to interpret the results and draw conclusions. This is an unique contribution merging AI and blockchain into communication protocol security. 

Similar focused research was done by M.Singh et al. It's also a merged methodology of blockchain and machine learning. But instead of deploying this research on a distributed network of IoD, the authors applied this technique on secure communication side. Ultimately, they claim the proposed methodology performed well enough to provide the required integrity of secured communication. The performance evaluation showed some scopes of improvement. Like before, computational power is a concern here. New concensus protocol needed to apply on IoD ecosystem to gain better results.

\subsubsection{Internet of Drones (IoD) security}
Basudeb Bera et al. proposed a private blockchain-based access control mechanism for unauthorized UAV detection and mitigation in Internet of Drones environment, which utilized smart contracts for authorization, authentication, and access control [2].The paper's motivation  is based on identifying  a research gap in the lack of a comprehensive access control mechanism that utilizes private blockchain for unauthorized UAV detection and mitigation in the IoD environment. The proposed methodology includes the use of smart contracts and consensus algorithms in a private blockchain network to establish access control rules, verify the authenticity of UAVs, and trigger mitigation actions for unauthorized UAVs.

With this base knowledge Aparna Kumari et al. present a taxonomy of blockchain-enabled software version for secure UAV networks, covering aspects such as trust, privacy, and security[7]. 

An interesting method which is different from other traditional methods, J. Daubert et al. presented "HoneyDrone". The authors aimed to detect and analyze attacks on UAV networks, and presents a novel approach for detecting and mitigating security threats in UAV networks by using a honeypot [6]. UAV honeypot is proposed to detect and analyze malicious attacks targeted at UAV networks, there's a significant difference between Honeypot and Block-chain based methods. 
The implementation of the HoneyDrone system is based on a Raspberry Pi and a Parrot AR Drone 2.0. The authors also discuss the different types of attacks that can be detected by the HoneyDrone, including wireless attacks, GPS spoofing, and denial-of-service attacks. But, this work has some limitations as well. The authors provides only limited evaluation results of the HoneyDrone honeypot, with experiments conducted in a controlled lab environment. More real-world evaluation is needed to fully understand the effectiveness of the approach. Another important aspect is ignored here is basically this model is designed for single UAV. This research can't be applied on IoD. 

To encounter that part, there are some IoT-blockchain based research work also found. Samir Dawaliby et al. proposed a blockchain-based IoT platform for autonomous drone operations management, which aimed to enable secure and efficient drone operations through blockchain-based identity and access management, and decentralized data sharing[9].

Another method mentioned by Basudeb Bera et al. Their research shows an extensive secure blockchain-based access control scheme in IoT-enabled Internet of Drones deployment, which utilized decentralized identity management and smart contract-based access control mechanisms to ensure secure and efficient UAV operations[10]. The scheme uses a permissioned blockchain network to store the access control policies and access logs of drones, operators, and ground stations. The access control policy is defined using smart contracts, which are executed on the blockchain network. The proposed scheme provides secure authentication and authorization of drones and operators, which ensures that only authorized drones and operators can access the resources of the IoD. This contribution made this work really unique in terms of merging both technologies. The evaluation is done by implementing a prototype using the Hyperledger Fabric blockchain framework. The evaluation results show that the proposed scheme can effectively manage the access control of drones and operators in a secure and efficient manner. 

Concepts of Internet of Drones IoD is becoming more and more popular and effective. The Internet of Drones (IoD) can be considered as a type of Flying Ad Hoc Network (FANET), which is a self-organizing network of drones that communicate with each other wirelessly to accomplish a specific task. C. Lin et al. identifies the security and privacy challenges faced by the Internet of Drones (IoD) and proposes solutions to mitigate these challenges [16].

\subsubsection{Deep Learning and Machine Learning based security}
Finally, some deep learning based novel diagram proposed by Joseanne Viana et al. This research proposed a deep attention recognition system for attack identification in 5G UAV scenarios, which employed deep learning techniques to enable accurate and efficient detection of attacks on UAV networks[11]. The proposed architecture uses an end-to-end approach that includes a deep neural network for feature extraction and attention mapping, followed by a fully connected network for classification of the attacks. The authors evaluate the proposed architecture using a publicly available dataset and compare its performance with several other state-of-the-art approaches. The results show that the proposed architecture outperforms the other approaches in terms of accuracy, sensitivity, and specificity. The paper also discusses the limitations of the proposed approach and suggests directions for future research, including the use of more sophisticated deep learning techniques and the exploration of other features for attack identification. One of the limitation mentioned in the paper is that it requires high computational resources, which may limit its applicability in real-time scenarios where low-latency processing is required. Also, The proposed model was evaluated under a specific set of attack scenarios, and its robustness to other types of attacks or variations in the environment is unknown. Under the consideration of hardware requirements, this proposed architecture relies on a specific hardware configuration that includes a high-end GPU, which may limit its scalability and practicality in real-world deployments.

The counterpart on ensuring the confidentiality and privacy is proposed by  Zhang et al. [12] . The authors proposed a novel multi-agent deep reinforcement learning (MADRL) framework for UAV-enabled secure communications approach based on multi-agent deep reinforcement learning.The proposed approach utilizes a combination of MADRL and cryptographic techniques to optimize the communication performance of UAVs and maintain security. The authors argue that traditional cryptographic methods might be insufficient in UAV networks due to limited computational power and communication bandwidth. The algorithm is trained using a simulation environment and tested in a real-world experiment using UAVs equipped with software-defined radios. The results show that the proposed approach can significantly improve the communication quality while maintaining a high level of security. The DRL-based algorithm outperforms traditional methods, such as fixed transmission power and random frequency selection. 

One limitation of the paper is that the proposed approach requires a significant amount of computational resources and may not be suitable for resource-constrained UAVs. Additionally, the proposed approach assumes a centralized control architecture, which may not be practical in large-scale UAV networks. Future work is directed on exploring the decentralized control architectures and alternative reinforcement learning methods that require less computational resources.

One of the major contribution on tackling DoS/DDoS attacks by Deep Learning has been done by Khan et al.[13]. By addressing an unique approach using  an experience-based deep learning algorithm, the authors showed a defense mechanism against denial of service attacks. This methodology didn't use any specific dataset to learn, rather it learned from the 'Experience'. This traditional method can be compared to reinforcement learning techniques, but there is a slight difference. In this method, the authors proposed event memories, so that the system know how the previous attacks happened. The results show that the proposed system is effective in detecting and mitigating these attacks and outperforms some existing approaches in terms of accuracy and efficiency. However, the paper does not provide a detailed analysis of the limitations and potential weaknesses of the proposed approach. Further research is needed to evaluate the scalability of the proposed system and its performance in more complex scenarios with a larger number of drones and nodes. Additionally, the authors have not compared their approach with some of the latest techniques for securing IoFT against DoS/DDoS attacks.

Comparing to this method, Reza Fotohi et al. provided an agent based self-protective defense against potential threats, unauthorized access and common attacks [5]. Though it's not a conventional machine/deep learning based model, but the type of the solution makes it more applicable for the shortcomings of previous methodology. The paper proposed an agent-based self-protective method that utilizes intelligent agents to enhance the security of communication between UAVs in UAV networks. The proposed methodology includes the deployment of intelligent agents on each UAV, which continuously monitor and analyze the communication environment for potential threats and attacks. The agents use rule-based and learning-based techniques to detect and mitigate security threats, such as jamming attacks, eavesdropping, and message alteration. The agents also collaborate with each other to exchange information and collectively make decisions to enhance the overall security of communication in the UAV network. The proposed methodology was implemented through a series of experiments and simulations. The experimental design involved setting up a realistic simulation environment that emulated the communication process between UAVs in a UAV network. The simulation scenarios were designed to test the performance, effectiveness, and robustness of the proposed methodology under different conditions, such as varying numbers of UAVs, communication distances, and types of attacks. Data was collected from the experiments and simulations to evaluate the performance, effectiveness, and robustness of the proposed methodology. Performance metrics, such as communication latency, packet loss, and system overhead, were measured and analyzed to assess the efficiency and effectiveness of the proposed methodology. Security metrics, such as threat detection rate, false positive rate, and system resilience, were also evaluated to assess the security of communication in the UAV network. Statistical analysis techniques, such as hypothesis testing and data visualization, were used to interpret the results and draw conclusions.

\subsubsection{Deployment scenario applications}
Different deployment scenarios can pose different security and privacy challenges in UAV networks. For example, a swarm of UAVs operating in a hostile environment may face different threats than a single UAV operating in a controlled environment. Alike this concept, few research has shown lights on \textit{'Edge Computing'}. One of edge-computing network method related to cyber defense in UAV-edge computing networks, including communication protocols, security measures, and cyber defense techniques.  is discussed by Hichem Sedjelmaci et al.[15]. The paper addresses the problem of cyber defense in UAV-edge computing networks and emphasizes on the challenges and vulnerabilities associated with UAV-edge computing networks, such as cyber-attacks and potential security breaches. The methodology includes a threat intelligence system, a security management system, a security analytics system, and a security orchestration system. The framework uses machine learning algorithms to detect and mitigate security threats, such as denial-of-service attacks, botnets, and malware.The experimental design involved setting up a realistic simulation environment that emulated the communication process between UAVs and the edge computing network. The simulation scenarios were designed to test the performance, effectiveness, and robustness of the proposed methodology under different conditions, such as varying network sizes, communication distances, and types of attacks. Data was collected from the experiments and simulations to evaluate the performance, effectiveness, and robustness of the proposed methodology. Performance metrics, such as detection rate, false positive rate, and system overhead, were measured and analyzed to assess the efficiency and effectiveness of the proposed methodology. Security metrics, such as threat detection rate, false positive rate, and system resilience, were also evaluated to assess the security of communication in the UAV-edge computing network. Statistical analysis techniques, such as hypothesis testing and data visualization, were used to interpret the results and draw conclusions.

Adding to that upper mentioned methodology an extra dimension added by Gope et al. on[18]. In this paper, the authors focused on edge-assisted IoD. The author addressed the problem of secure key agreement in edge-assisted Internet of Drones (IoD) networks and highlighted the importance of securing communication between drones and the edge computing infrastructure to prevent unauthorized access, man-in-the-middle attacks, and other security threats. The need for an efficient, privacy-preserving authenticated key agreement scheme for IoD networks is emphasized. The paper proposed an efficient, privacy-preserving authenticated key agreement scheme that utilizes elliptic curve cryptography (ECC) to secure communication between drones and the edge computing infrastructure. The proposed scheme involves three phases: initial setup, key agreement, and key confirmation. In the initial setup phase, the drones and the edge computing infrastructure exchange their public keys and a shared secret key is generated. In the key agreement phase, the drones and the edge computing infrastructure use the shared secret key to generate a session key for secure communication. In the key confirmation phase, the session key is verified using a zero-knowledge proof protocol. The proposed scheme ensures privacy preservation by using elliptic curve Diffie-Hellman (ECDH) key exchange and digital signature techniques. The proposed methodology was implemented through a series of simulations using the NS-3 network simulator. The experimental design involved setting up a realistic simulation environment that emulated the communication process between drones and the edge computing infrastructure in an IoD network. The simulation scenarios were designed to test the performance, effectiveness, and robustness of the proposed methodology under different conditions, such as varying numbers of drones, communication distances, and types of attacks. This was validated through comparison with existing key agreement schemes and through the use of mathematical models and simulations. 

On the deployment side, another application is free space communication via UAV and ground stations. This is the backbone of 5G/6G communication. Q.Huang et al. mentioned a method to use data fragmentation and multipath transmission technology to ensure secure communication between UAVs and ground stations using free-space optical communication[14]. The paper addresses the problem of securing  free-space optical communication (FSOC)against eavesdropping attacks and ensuring reliable communication in the presence of atmospheric turbulence. The paper highlights the challenges associated with achieving secure and reliable communication in conventional FSOC systems and proposes a novel methodology to address these challenges. The paper proposes a secure FSOC system that utilizes data fragmentation and multipath transmission technologies to enhance the security and reliability of communication. The proposed methodology involves the fragmentation of data into multiple smaller packets, which are transmitted over multiple paths simultaneously. The fragments are reassembled at the receiver end, ensuring reliable delivery of data even in the presence of atmospheric turbulence. The use of multiple paths also enhances the security of communication by reducing the likelihood of eavesdropping attacks. The proposed methodology was evaluated through a series of experiments conducted in a laboratory environment. The experimental design involved setting up a realistic simulation environment that emulated the communication process between the transmitter and receiver in a FSOC system. The simulation scenarios were designed to test the performance, effectiveness, and robustness of the proposed methodology under different conditions, such as varying atmospheric turbulence levels and numbers of fragments and paths. he paper acknowledges several limitations associated with the proposed methodology. The main limitation is that the proposed methodology may not be suitable for long-distance communication due to the increased susceptibility to atmospheric turbulence and the need for multiple paths. The paper also notes that the proposed methodology may require additional computational resources to implement the data fragmentation and reassembly processes, which may increase the overall system complexity and cost.

\section{Findings}

From the survey, I found a detailed threat issues and control mechanism in the field of UAV-communication applications. Following the ISO 27001 standard, the papers introduced various types of threats related to the topic. In the section "Methods-Threat Models", the scenario based threat modelling is discussed. An ISO 27001-27005 ISMS basically holds the accountability of confidentiality, Integrity and availability (CIA) of information security. In the Methods section, how these paper follow and focus the CIA has been mentioned.

The security models sections follows the ISO 27001 risk assessment, management and control. Based on the vulnerability assessment, the authors brought up different methods like, block-chain, AI, IoD, machine learning, edge computing etc. Newer technologies demand more robust and sophisticated solutions which is a part of ISMS requirements. 5G and 6G communication protocol follows the EEE 802.11 wireless routing protocols standard. Khan et al. [13] mentioned the integrity that their research approved, and the proposed methodology was built for that standard. Similarly. the deep learning and blockchain based solutions are mainly focused on reviewing and improving ISMS system as these are dataset oriented technologies. Once there is any vulnerability, there's more scope for improvement in terms of learning. Deep learning plays this role inherently, some of the methodologies are mentioned in details in the Methods part.

\section{Conclusion and Future Work}
The study concludes with some recent technologies which can be deployed in many application in future. The solutions are not limited to, but addressed here for example, block-chain, deep learning, machine learning, cryptographic solutions. Some deployed scenarios also discussed like secure free-space communication, edge computing security. In general, Deep Learning methods perform a little quicker comparing block-chain and other methods. But, if we take the consideration of security then block-chain and cryptographic solutions are more robust. There is a overall common drawback can be seen in these papers which is computational complexity. To reduce vulnerability, the authors came up with various solutions which mostly beat the previous results in terms of security concern. But, this process takes a lot more computational power and spaces. In future when the technologies will be there, some authors directed to the development of hardware management issues. This is a core requirement of ISO standard ISMS.

\bibliographystyle{ieeetr}
{
	\bibliography{Ref}
}

\end{document}